\documentclass[showpacs,preprintnumbers,amsmath,amssymb,nofootinbib]{revtex4}
\usepackage{epsfig,ulem}
\usepackage{amsmath}
\usepackage{amssymb}
\usepackage{amsfonts}
\usepackage{graphicx}

\newcommand{\si}{\sigma}
\newcommand{\sq}{\tilde{q}}
\newcommand{\sqb}{\bar{\tilde{q}}}
\newcommand{\gl}{\tilde{g}}

\newcommand{\NLO}{\mathrm{NLO}}

\newcommand{\TeV}{\unskip\,\mathrm{TeV}}

\newcommand{\met}{E_{T, {\rm miss}}}

\def\nn{\nonumber}

\def\text{\textstyle}

\def\bc{\begin{center}}
\def\ec{\end{center}}
\def\bi{\begin{itemize}}
\def\ei{\end{itemize}}

\begin{document}
    
\title{Improved squark and gluino mass limits from searches for supersymmetry at hadron colliders}

\author{Wim Beenakker}
\affiliation{Theoretical High Energy Physics, IMAPP
Mailbox 79, P.O. Box 9010, NL-6500 GL Nijmegen, The Netherlands}
  
\author{Silja Brensing}
\affiliation{DESY, Theory Group, Notkestrasse 85, D-22603 Hamburg, Germany}

\author{Monica D'Onofrio}
\affiliation{University of Liverpool, Oliver Lodge Laboratory, P.O. Box 147, Oxford Street, Liverpool L69 3BX, United
Kingdom}

\author{Michael Kr\"amer}
\affiliation{Institute for Theoretical Particle Physics and Cosmology, RWTH Aachen University,
  D-52056 Aachen, Germany}

\author{Anna Kulesza}
\affiliation{Institute for Theoretical Particle Physics and Cosmology, RWTH Aachen University, 
  D-52056 Aachen, Germany}

\author{Eric Laenen}
\affiliation{ITFA, University of Amsterdam, Science Park 904, 1090 GL Amsterdam, \\
  ITF, Utrecht University, Leuvenlaan 4, 3584 CE Utrecht\\
  Nikhef, Science Park 105, 1098 XG Amsterdam, The
  Netherlands}

\author{Mario Martinez}
\affiliation{ICREA and Institut de F\'{i}sica dÕAltes Energies, IFAE, Edifici Cn, Universitat Aut\`{o}noma de Barcelona, ES - 08193 Bellaterra
(Barcelona), Spain}

\author{Irene Niessen}
\affiliation{{}Theoretical High Energy Physics, IMAPP
Mailbox 79, P.O. Box 9010, NL-6500 GL Nijmegen, The Netherlands}

\begin{abstract}
\noindent
Squarks and gluinos have been searched for at hadron colliders in
events with multiple jets and missing transverse energy. No excess has
been observed to date, and from a comparison of experimental cross
section limits and theoretical cross section predictions one can
deduce lower bounds on the squark and gluino masses. We present an
improved analysis of squark and gluino mass bounds which is based on
state-of-the-art cross section calculations including the summation of
large threshold corrections. For illustration, we consider
experimental data obtained by the CDF collaboration at the Fermilab
Tevatron and discuss the impact of the improved cross section
predictions on the squark and gluino mass limits.  
\end{abstract}

\maketitle

\section{Introduction}
\label{sec:intro}
The search for supersymmetry (SUSY)~\cite{Golfand:1971iw, Wess:1974tw}
is one of the central tasks at the currently operating hadron
colliders Tevatron and LHC. In most SUSY models the squarks ($\tilde
q$) and gluinos ($\tilde g$), the coloured supersymmetric particles (sparticles),
are produced copiously in hadronic collisions.  We consider the
minimal supersymmetric extension of the Standard Model
(MSSM)~\cite{Nilles:1983ge, Haber:1984rc} where, as a consequence of
R-parity conservation, squarks and gluinos are produced in pairs and
where the lightest supersymmetric particle (LSP) is stable. For large
regions of parameter space the LSP is also neutral and weakly
interacting, and provides a viable dark matter candidate.  The expected
MSSM signature at colliders is thus characterized by multiple jets
from cascade decays of squarks and gluinos and by large missing
transverse energy $\met$ from the two LSPs at the end of
each decay chain.

Searches for squarks and gluinos at the proton--antiproton collider
Tevatron with a centre-of-mass energy of $\sqrt{S}=1.96$~TeV have
placed lower limits on squark and gluino masses in the range of
300-400~GeV~\cite{:2007ww,Aaltonen:2008rv} in a minimal supergravity
(mSUGRA) scenario. The proton--proton collider LHC currently operating
at $\sqrt{S}=7$~TeV has already significantly extended the mSUGRA
mass limits to values near 800~GeV~\cite{Khachatryan:2011tk,
  daCosta:2011hh}. The mass bounds are deduced from the comparison of
experimental cross section exclusion limits and theoretical cross
section predictions. So far, the interpretation of the experimental
search results have relied on calculations of the squark and gluino
cross sections at next-to-leading order (NLO) in
SUSY-QCD~\cite{Beenakker:1994an, Beenakker:1995fp, Beenakker:1996ch,
  Beenakker:1997ut}. Recently, the accuracy of these cross section
predictions has been improved further by the resummation of soft-gluon
emission to all orders in perturbation theory~\cite{Kulesza:2008jb,
  Kulesza:2009kq, Beenakker:2009ha, Beenakker:2010nq, Beenakker:2011fu}. 
  The inclusion
of soft-gluon resummation leads to a significant reduction of the
unphysical renormalization and factorization scale dependence and, in
general, increases the NLO cross sections.

In this work we employ state-of-the-art SUSY-QCD cross section
predictions including next-to-leading-logarithmic (NLL) soft-gluon resummation~\cite{Kulesza:2008jb,
  Kulesza:2009kq, Beenakker:2009ha, Beenakker:2010nq, Beenakker:2011fu} 
  to derive improved mass bounds on squarks and gluinos from the analysis of
recent CDF data \cite{Aaltonen:2008rv}. 

The paper is structured as follows. In section~\ref{se:resummation} we
briefly summarize the status of cross section calculations for squark
and gluino hadroproduction and discuss the improvement obtained from
soft-gluon resummation.  The experimental CDF analysis is reviewed in
section~\ref{se:experiment}. The interpretation of the cross section
limits in terms of squark and gluino mass bounds is discussed in
section~\ref{se:massbounds}, where we also address the treatment of
theoretical uncertainties.  We conclude in
section~\ref{se:conclusion}.

\section{Squark and gluino cross sections}
\label{se:resummation}

In the MSSM with R-parity conservation, squarks and gluinos are produced in pairs, 
\begin{equation}
  pp/p\bar{p} \;\to\; \tilde{q}\bar{\tilde{q}}, \tilde{q}\tilde{q}, \tilde{q}\tilde{g}, \tilde{g}\tilde{g} + X\,.
\label{eq:processes}
\end{equation}
In Eq.~(\ref{eq:processes}) and throughout the rest of this paper we
suppress the chiralities of the squarks $\tilde{q} =(\tilde{q}_{L},
\tilde{q}_{R})$ and do not explicitly state the charge-conjugated
processes.  We include squarks $\tilde{q}$ of any flavour except for
top squarks. The production of top squarks~\cite{Beenakker:1997ut}
needs to be considered separately because of potentially large mixing
effects and mass splitting in the stop sector~\cite{Ellis:1983ed}.

The cross sections for the squark and gluino pair-production processes
(\ref{eq:processes}) have been known at next-to-leading order in
SUSY-QCD~\cite{Beenakker:1994an, Beenakker:1995fp, Beenakker:1996ch,
  Beenakker:1997ut} for some time. Electroweak corrections to the
${\cal O} (\alpha_{\rm s}^2)$ tree-level
production~\cite{Hollik:2007wf, Beccaria:2008mi, Hollik:2008yi,
  Hollik:2008vm, Mirabella:2009ap} and the electroweak Born production
channels of ${\cal O} (\alpha\alpha_{\rm s})$ and ${\cal O}
(\alpha^2)$~\cite{Alan:2007rp, Bornhauser:2007bf} are significant for
the pair production of SU(2)-doublet squarks $\tilde{q}_L$ and at
large invariant masses in general, but they are moderate for total
cross sections summed over all squark species.

The NLO SUSY-QCD corrections to squark and gluino hadroproduction
reduce the renormalization- and factorization-scale dependence of the
predictions.  In general the NLO contributions also significantly
increase the cross section with respect to the leading-order (LO)
approximation~\cite{Kane:1982hw, Harrison:1982yi, Dawson:1983fw} if
the renormalization and factorization scales are chosen close to the
average mass of the pair-produced sparticles. A significant part of
the NLO corrections can be attributed to the threshold region, where
the partonic centre-of-mass energy is close to the kinematic
production threshold. In this region the NLO corrections are dominated
by the contributions due to soft gluon emission off the coloured
particles in the initial and final state and by the Coulomb
corrections due to the exchange of gluons between the massive
sparticles in the final state.  The soft-gluon corrections can be
taken into account to all orders in perturbation theory by means of
threshold resummation techniques.

Recently, such threshold resummation has been performed for all squark
and gluino production processes in Eq.\,(\ref{eq:processes}) at
NLL accuracy~\cite{Kulesza:2008jb,
  Kulesza:2009kq, Beenakker:2009ha, Beenakker:2010nq}. For
squark-antisquark production, NLL resummation has also been addressed
in the framework of effective field theories \cite{Beneke:2009rj,
  Beneke:2009nr}, and the dominant next-to-next-to-leading order
(NNLO) correction coming from the resummed cross section at
next-to-next-to-leading-logarithmic (NNLL) level has been derived in
Ref.\,\cite{Langenfeld:2009eg, Langenfeld:2010vu}.  Moreover, a formalism allowing for
the resummation of soft and Coulomb gluons in the production of
coloured sparticles has been presented in Refs.\,\cite{Beneke:2009rj,
  Beneke:2009nr}, and bound state effects have been studied for
gluino-pair production in Ref.~\cite{Hagiwara:2009hq}.

In this work we will employ the state-of-the-art SUSY-QCD predictions
at NLO+NLL for squark and gluino production to derive mass bounds from
the comparison with experimental exclusion limits. In the remainder of
this section we briefly review the calculation of the NLO+NLL cross
section presented in \cite{Kulesza:2008jb, Kulesza:2009kq,
  Beenakker:2010nq, Beenakker:2009ha, Beenakker:2011fu} and present a few illustrative
numerical results.

The hadronic threshold for inclusive production of two final-state
particles with masses $m_k$ and $m_l$ corresponds to a hadronic
center-of-mass energy squared that is equal to $S=(m_k+m_l)^2$.  Thus
we define the threshold variable $\rho$, measuring the distance from
threshold in terms of energy fraction, as
\begin{equation}
  \label{eq:1}
\rho \;=\; \frac{(m_k+m_l)^2}{S}\,.  
\end{equation}
Our results are based on the following expression for the NLL-resummed
cross section matched to the exact NLO calculation of
Refs.\,\cite{Beenakker:1994an, Beenakker:1995fp, Beenakker:1996ch}
\begin{eqnarray}
\label{eq:14}
\si^{\rm (NLO+NLL)}_{h_1 h_2 \to kl}\bigl(\rho, \{m^2\},\mu^2\bigr) 
  &=& \si^{\rm (NLO)}_{h_1 h_2 \to kl}\bigl(\rho, \{m^2\},\mu^2\bigr)\nn
          \\[1mm]
   &&  \hspace*{-30mm}+\, \frac{1}{2 \pi i} \sum_{i,j=q,\bar{q},g}\, \int_\mathrm{CT}\,dN\,\rho^{-N}\,
       \tilde f_{i/h_1}(N+1,\mu^2)\,\tilde f_{j/h_{2}}(N+1,\mu^2) \nn\\[0mm]
   && \hspace*{-20mm} \times\,
       \left[\tilde\si^{\rm(res)}_{ij\to kl}\bigl(N,\{m^2\},\mu^2\bigr)
             \,-\, \tilde\si^{\rm(res)}_{ij\to kl}\bigl(N,\{m^2\},\mu^2\bigr)
       {\left.\right|}_{\scriptscriptstyle({\NLO})}\, \right]\,,
\end{eqnarray}
where the last term in the square brackets denotes the NLL resummed
expression expanded to NLO.  The initial state hadrons are
denoted generically as $h_1$ and $h_2$, while $\mu$ is the common
renormalization and factorization scale. The resummation is performed after
taking a Mellin transform (indicated by a tilde) of the cross section,
\begin{equation}
  \label{eq:10}
  \tilde\si_{h_1 h_2 \to kl}\bigl(N, \{m^2\}\bigr) 
 \equiv \int_0^1 d\rho\;\rho^{N-1}\;
           \si_{h_1 h_2\to kl}\bigl(\rho,\{ m^2\}\bigr) \,.
\end{equation}
To evaluate the contour CT of the inverse Mellin transform in
Eq.~(\ref{eq:14})  we adopt the ``minimal prescription'' of
Ref.~\cite{Catani:1996yz}. The NLL resummed cross section in
Eq.~(\ref{eq:14}) reads
\begin{multline}
  \label{eq:12}
  \tilde{\sigma}^{\rm (res)} _{ij\rightarrow kl}\bigl(N,\{m^2\},\mu^2\bigr) 
= \sum_{I}\,
      \tilde\sigma^{(0)}_{ij\rightarrow kl,I}\bigl(N,\{m^2\},\mu^2\bigr)\, 
      C_{ij \rightarrow kl, I}\bigl(N,\{m^2\},\mu^2\bigr) \\[1mm]
   \times\,\Delta_i (N+1,Q^2,\mu^2)\,\Delta_j (N+1,Q^2,\mu^2)\,
     \Delta^{\rm (s)}_{ij\rightarrow
       kl,I}\bigl(N+1,Q^2,\mu^2\bigr)\,,
\end{multline}
where we have introduced the hard scale
$Q^2=(m_k+m_l)^2$. The colour-decomposed leading-order cross sections
in Mellin-moment space are denoted by $\tilde\sigma^{(0)}_{ij\to
kl,I}$ with $I$ labeling the different possible colour structures.
The expressions
for these, both in moment and in momentum space, can be found in
\cite{Kulesza:2009kq,Beenakker:2009ha}.  The perturbative functions
$C_{ij \rightarrow kl, I}$ contain information about hard
contributions beyond leading order.  This information is only relevant
beyond NLL accuracy and therefore we keep $C_{ij \rightarrow kl,I} =1
$ in our calculations. The functions $\Delta_{i}$ and $\Delta_{j}$ sum
the effects of the (soft-)collinear radiation from the incoming
partons.  They are process-independent and do not depend on the colour
structures.  They contain the leading logarithmic dependence, as well
as part of the subleading logarithmic behaviour, and are listed e.g.\
in Ref.~\cite{Kulesza:2009kq}. The resummation of the soft-gluon
contributions, which does depend on the colour structures in which the
final state SUSY particle pairs can be produced, contributes at the
NLL level and is summarized by the factor
\begin{equation}
  \Delta_{_{ij\rightarrow
       kl,I}}^{\rm (s)}\bigl(Q/(N\mu),\mu^2\bigr) 
  \;=\; \exp\Big[\int_{\mu}^{Q/N}\frac{dq}{q}\,\frac{\alpha_{\rm s}(q)}{\pi}
                 \,D_{I} \,\Big]\,.
\label{eq:2}
\end{equation}
The one-loop coefficients $D_{I}$ are derived and listed in
\cite{Kulesza:2009kq, Beenakker:2009ha}.

We now present a few selected numerical results to illustrate the
impact of the NLL resummation on squark and gluino production at the
Tevatron.  We compare the LO, NLO and NLO+NLL matched results and
discuss the theoretical uncertainty due to the choice of
renormalization and factorization scales, and the parametrization of the
parton distribution functions (PDFs).  The NLO cross sections have been calculated in
Refs.~\cite{Beenakker:1994an, Beenakker:1995fp, Beenakker:1996ch,
  Beenakker:1997ut} and are available in the form of the public computer
code {\tt Prospino}\,\cite{prospino}.  The $\overline{\rm MS}$-scheme
with five active flavours is used to define the QCD coupling
$\alpha_{\rm s}$ and the parton distribution functions at NLO. The
masses of the squarks and gluinos are renormalized in the on-shell
scheme, and the SUSY particles are decoupled from the running of
$\alpha_{\rm s}$ and the PDFs. 

\begin{figure}
\hspace{-0.2cm}
\begin{tabular}{ll}
\epsfig{file=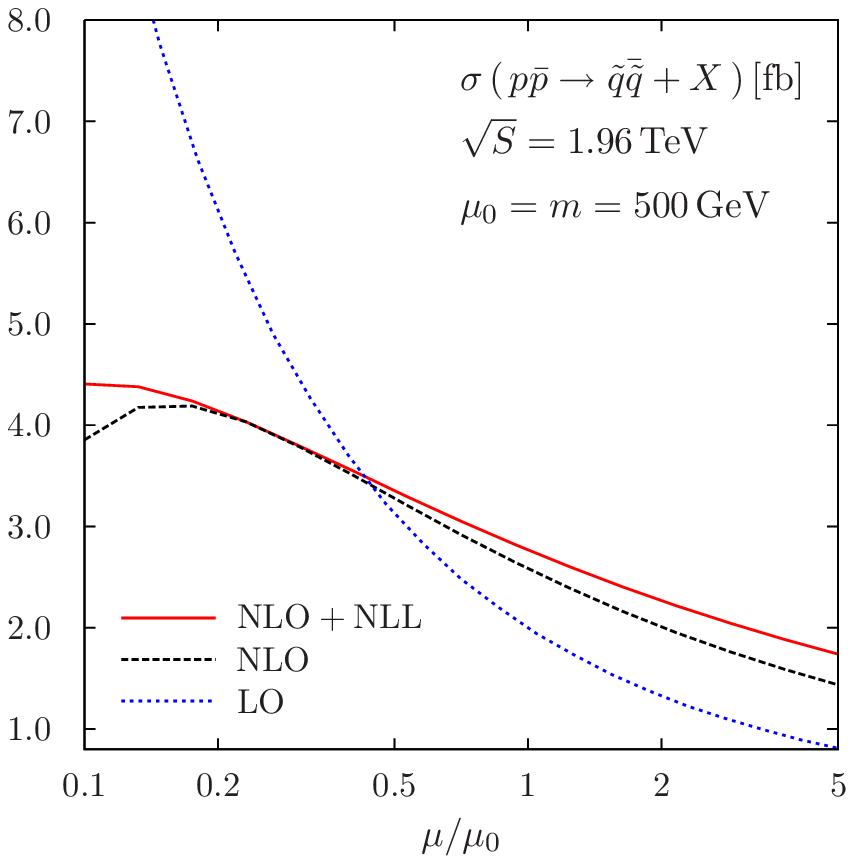, width=0.465\columnwidth}& 
\;\;\;\epsfig{file=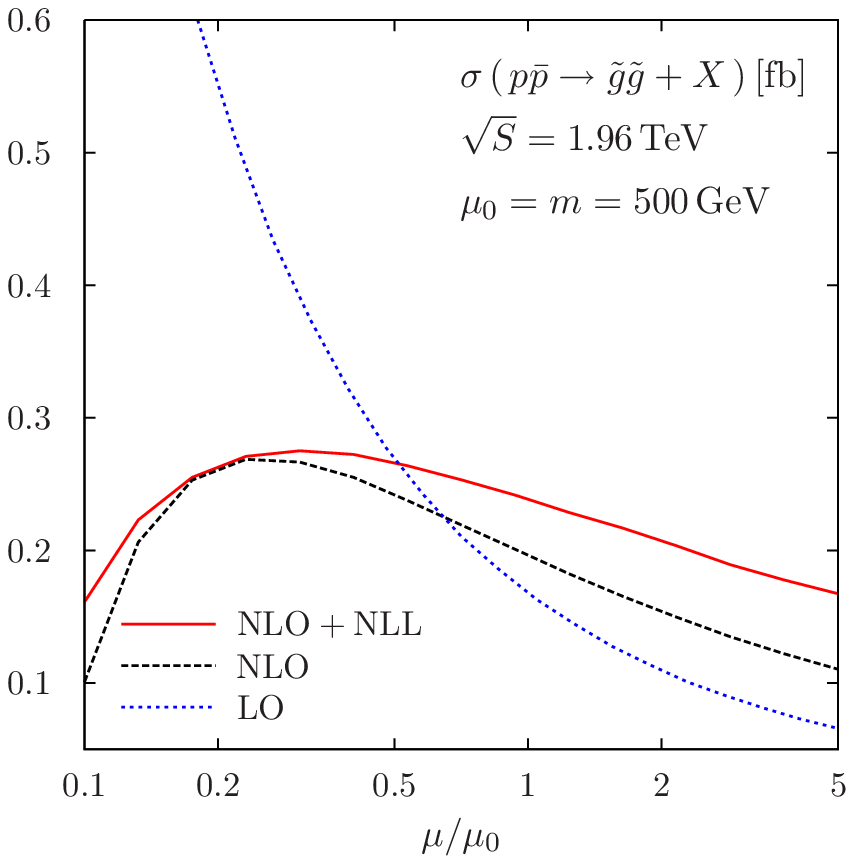, width=0.465\columnwidth }\\
\epsfig{file=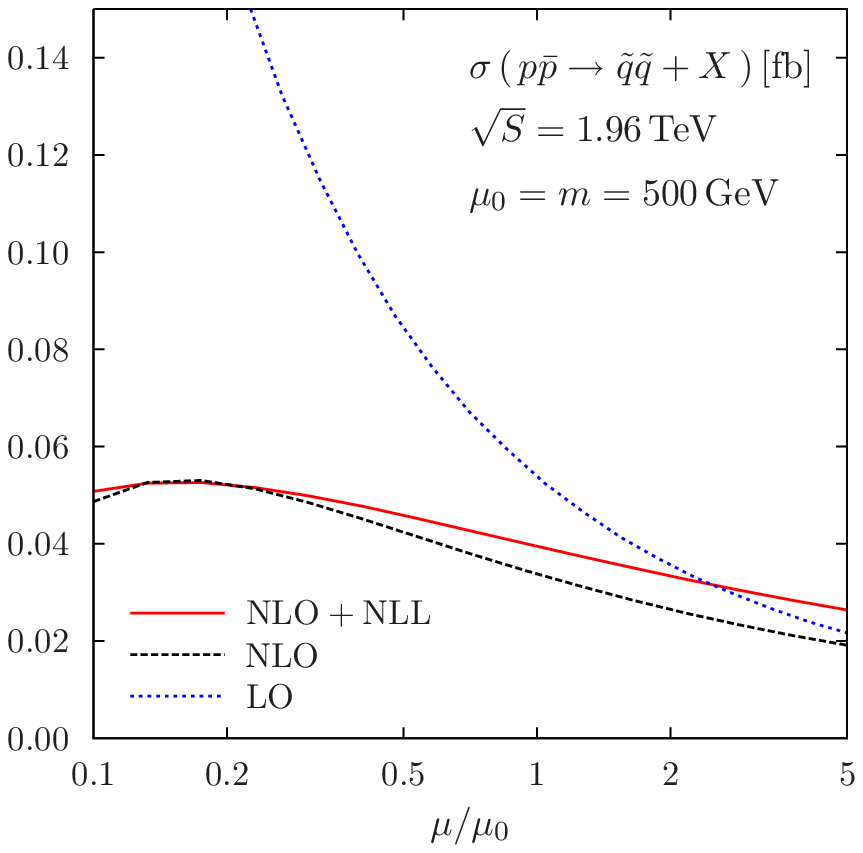, width=0.465\columnwidth}& 
\;\;\;\epsfig{file=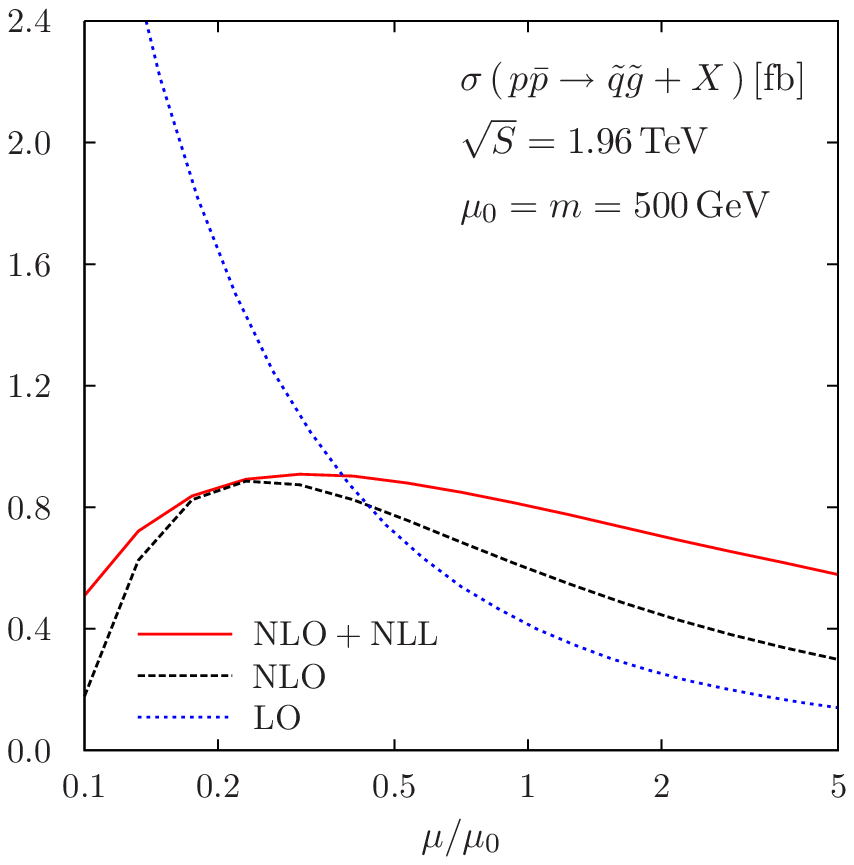, width=0.465\columnwidth }\\
\end{tabular}
\caption{The scale dependence of the LO, NLO and NLO+NLL cross
  sections for the four different squark and gluino production
  processes at the Tevatron. The squark and gluino masses have been
  set to $m_{\sq}=m_{\gl}=m=500$~GeV.  The MSTW-2008\,\cite{MSTW} PDF
  has been adopted.}
\label{fig:scale}
\end{figure}

We first discuss the scale dependence of the SUSY-QCD cross-section
prediction for the four different production processes $p\bar{p} \to
\sq\sqb, \sq\sq, \sq\gl, \gl\gl + X$ at the Tevatron. For convenience
we define the average mass of the final-state sparticle pair $m = (m_k
+ m_l)/2$, which reduces to the squark and gluino mass for
$\tilde{q}\bar{\tilde{q}}$, $\tilde{q}\tilde{q}$ and $\tilde{g}\tilde{g}$
final states, respectively. The renormalization and factorization
scales are taken to be equal.  Figure~\ref{fig:scale} compares the scale
dependence in LO, NLO and NLO+NLL for $m_{\sq}=m_{\gl}=m=500$~GeV.
The scale $\mu$ is varied around the standard scale choice $\mu_0 = m$
from $\mu=\mu_0/10$ up to $\mu=5 \,\mu_0$.  Note that the LO
predictions are obtained with LO PDFs and the corresponding LO values
for $\alpha_{\rm s}$. We observe the anticipated strong reduction of
the scale dependence when going from LO to NLO, and a further
significant improvement when the resummation of threshold logarithms
is included, in particular for $\gl\gl$ and $\sq\gl$ production. 

\begin{figure}
\hspace{-1cm}
\epsfig{file=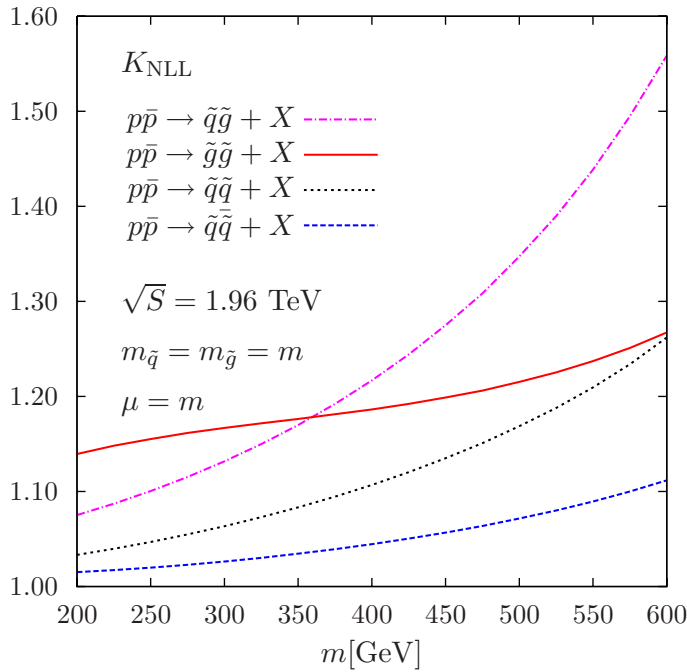, width=0.5\columnwidth}
\caption{The NLL $K$-factor $K_{\rm NLL} = \sigma_{\rm
    NLO+NLL}/\sigma_{\rm NLO}$ for squark and gluino pair-production
  processes at the Tevatron as a function of the average sparticle
  mass $m$. The MSTW-2008\,\cite{MSTW} PDF has been adopted.}
\label{fig:k_tev}
\end{figure}

At the standard choice of scale $\mu=\mu_0=m$ the cross-section
predictions are in general enhanced by soft-gluon resummation.  The
NLL $K$-factor $K_{\rm NLL} \equiv \sigma_{\rm 
  NLO+NLL}/\sigma_{\rm NLO}$ at the Tevatron is displayed in
Figure~\ref{fig:k_tev} for squark and gluino masses in the range between
200~GeV and 600~GeV.
We show results for equal squark and gluino masses, $K$-factors for
different mass ratios $m_{\tilde{g}}/m_{\tilde{q}}$ can be found in
Ref.\,\cite{Beenakker:2009ha}.  The soft-gluon corrections are
moderate for $\sq\sqb$ production, but very significant for
$\gl\gl\,$, $\sq\sq$ and $\sq\gl$ final states, respectively. Because
of the increasing importance of the threshold region, the corrections
become in general larger for increasing sparticle masses.  The large
effect of soft-gluon resummation for $\tilde{q}\tilde{g}$ and
$\tilde{g}\tilde{g}$ production can be mostly attributed to the
importance of gluon initial states for these processes. Furthermore,
the presence of gluinos in the final state results in an enhancement of
the NLL contributions since the Casimir invariants which enter the
resummation formulae are larger than for processes involving only
squarks.  The substantial value of $K_{\rm NLL}$ for $\sq\sq$
production at the Tevatron is a consequence of the behaviour of the
corresponding NLO corrections, which strongly decrease with increasing
squark mass \cite{Beenakker:1996ch}.

Representative values for the LO, NLO and NLO+NLL cross sections are
collected in Table~\ref{tab:tev}. The CDF analysis \cite{Aaltonen:2008rv} 
was based on cross section predictions obtained 
with the CTEQ6.1\,\cite{Stump:2003yu} PDF set. In Table~\ref{tab:tev} we thus 
compare 
predictions for the old CTEQ6.1 PDF and for the more recent 
CTEQ6.6\,\cite{Nadolsky:2008zw} PDF, which we use to derive 
the improved cross section limits presented in this paper. 

\renewcommand{\arraystretch}{1.5}
\begin{table}
\begin{footnotesize}
 \begin{tabular}{c|c|c||c|c}
 \multicolumn{5}{c}{\normalsize $p\bar{p} \;\to\; \tilde{q}\bar{\tilde{q}} + \tilde{q}\tilde{q} + \tilde{q}\tilde{g} + \tilde{g}\tilde{g} + X$ at $\sqrt{S}=1.96\TeV$}
 \\[1mm] \hline 
 \multicolumn{1}{c}{} & \multicolumn{2}{c||}{CTEQ6.1} & \multicolumn{2}{c}{CTEQ6.6} \\[1mm] \hline
$m_{\tilde{q}}/m_{\tilde{g}} \; [\mathrm{GeV}]$ & 400/400 & 460/300 & 400/400 & 460/300 \\  \hline \hline
$(\sigma\pm\Delta\sigma_{\mu})_\mathrm{LO} \; [\mathrm{pb}]$ & $(5.39 ^{+2.66}_{-1.66}) \times 10^{-2}$  &  $(1.51 ^{+0.71}_{-0.44}) \times 10^{-1} $  & $(5.39 ^{+2.66}_{-1.66}) \times 10^{-2}$ &  $(1.51 ^{+0.71}_{-0.44}) \times 10^{-1} $ \\  \hline 
$(\sigma\pm\Delta\sigma_{\mu})_\mathrm{NLO} \; [\mathrm{pb}]$ & $(8.83 ^{+1.96}_{-1.82}) \times 10^{-2}$ & $(3.41^{+0.90}_{-0.76}) \times 10^{-1}$ & $(8.49 ^{+1.85}_{-1.73}) \times 10^{-2}$ & $(3.16 ^{+0.81}_{-0.69}) \times 10^{-1} $  \\  \hline
$(\sigma\pm\Delta\sigma_{\mu})_\mathrm{NLO+NLL} \; [\mathrm{pb}]$ & $(9.71 ^{+1.44}_{-1.45})\times 10^{-2}$ & $(3.77 ^{+0.68}_{-0.60})\times 10^{-1}$ & $(9.30^{+1.37}_{-1.38})\times 10^{-2}$ & $(3.47 ^{+0.62}_{-0.54})\times 10^{-1}$ \\  \hline
$\Delta\mathrm{pdf}_\mathrm{NLO} \; [\%]$ & ${}^{+26}_{-14}$ & ${}^{+32}_{-18}$  & ${}^{+16}_{-10}$  & ${}^{+19}_{-13}$ \\  \hline 
$\mathrm{K}_{\mathrm{NLO}}$ & 1.64 & 2.26 & 1.58 & 2.09  \\ \hline 
$\mathrm{K}_{\mathrm{NLL}}$ & 1.10 & 1.11 & 1.10 & 1.10  \\
 \end{tabular}

\caption{The LO, NLO and NLO+NLL cross
  sections for inclusive sparticle pair production 
  $p\bar{p} \;\to\; \tilde{q}\bar{\tilde{q}} + \tilde{q}\tilde{q} + \tilde{q}\tilde{g} + \tilde{g}\tilde{g} + X$  
  at the Tevatron
  ($\sqrt{S}$=1.96 TeV), including errors due to scale variation
  ($\Delta\sigma_{\mu}$) in the range $m/2 \le \mu \le
  2m$. Results are shown for two PDF parametrizations
  (CTEQ6.1\,\cite{Stump:2003yu}, CTEQ6.6\,\cite{Nadolsky:2008zw}) with 
  the corresponding 68\% C.L.\ PDF error
  estimates.}
\label{tab:tev}
\end{footnotesize}
\end{table}

As discussed before,
we observe an increase of the cross-section prediction near the central scale 
when going from LO to NLO and a further enhancement when NLL threshold 
resummation is included. The scale dependence in the range 
$m/2 \le \mu \le 2m$ is reduced from about 
$\pm 25\%$ at NLO to about $\pm 15\%$ at NLO+NLL. The estimated PDF uncertainty
is approximately 15\%. The cross section is reduced when going from the 
old CTEQ6.1\,\cite{Stump:2003yu} to the more modern  CTEQ6.6\,\cite{Nadolsky:2008zw}
PDF set.


\section{Experimental searches for squarks and gluinos}
\label{se:experiment}

Experimental searches for inclusive squark and gluino production are
based on the study of events  with  multiple jets of hadrons 
and large $\met$ in the final state. In the CDF study~\cite{Aaltonen:2008rv} at the Tevatron,
the SUSY samples were generated using the ISASUGRA implementation in PYTHIA\,\cite{Sjostrand:2000wi}
and normalized
according to  NLO cross sections as determined using {\tt Prospino}~2.0  with CTEQ6.1M PDFs
and the renormalization and factorization 
scales set equal to the average mass of the sparticles produced in the hard interaction.  The uncertainties 
in the NLO predictions due to PDFs varied between 10$\%$ and $25\%$ across the region of the 
squark/gluino mass plane considered, and 
variations by a factor two of the renormalization and factorization scales changed the theoretical predictions by 
20$\%$ to $25\%$.  
Final states with different inclusive 
jet multiplicity were considered to optimize the sensitivity across the 
quark/gluino mass plane. In the scenario with squark masses significantly 
larger than the gluino mass at least four jets in the final state are expected, while 
for a gluino mass much larger than the squark masses dijet configurations dominate. 
In the CDF analysis, three parallel analyses in  
event topologies with large $\met$ and at least four, three, and two jets in the final state were carried out.
In each case the event selection criteria were optimized such as to
maximize the sensitivity to the SUSY signal. 
As an example, for degenerate squarks and gluinos with masses 
around 400~GeV, $S/\sqrt{B} = 6$~\footnote{Here $S$ and $B$ denote the number of signal and SM background 
events, respectively.} was obtained, with  a signal efficiency, defined as the fraction of signal events 
passing the selection criteria, of about 12$\%$. Additional details
on the CDF data analysis can be found in~\cite{Lorenzo:2010th,cdfweb}. 
In this paper, CDF results are revisited with improved NLO+NLL theoretical 
predictions, leading to reduced scale uncertainties, and updated CTEQ6.6 PDF sets.

\begin{figure}
\hspace{-0.2cm}
\begin{tabular}{ll}
\epsfig{file=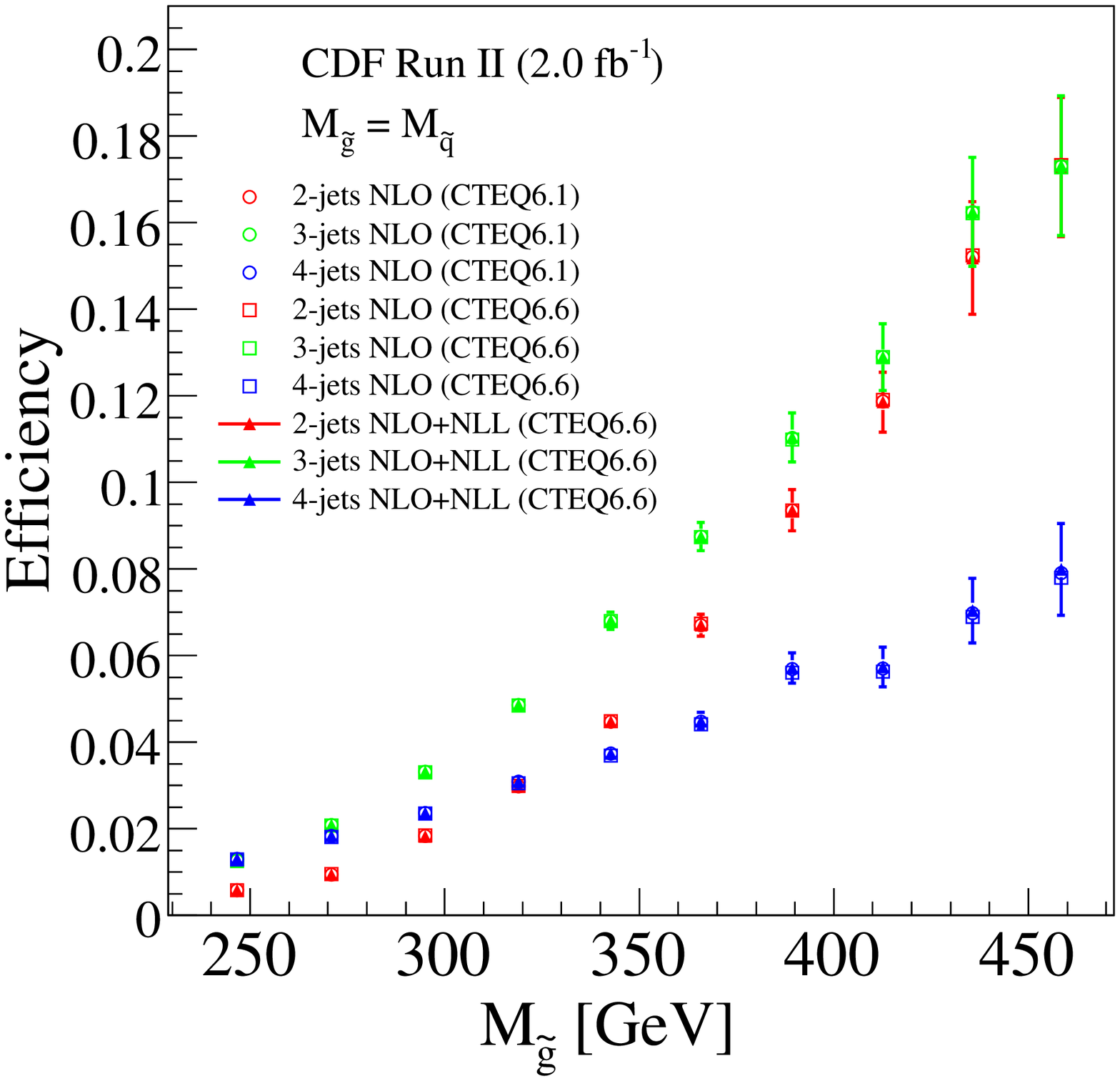, width=0.465\columnwidth}& 
\;\;\;\epsfig{file=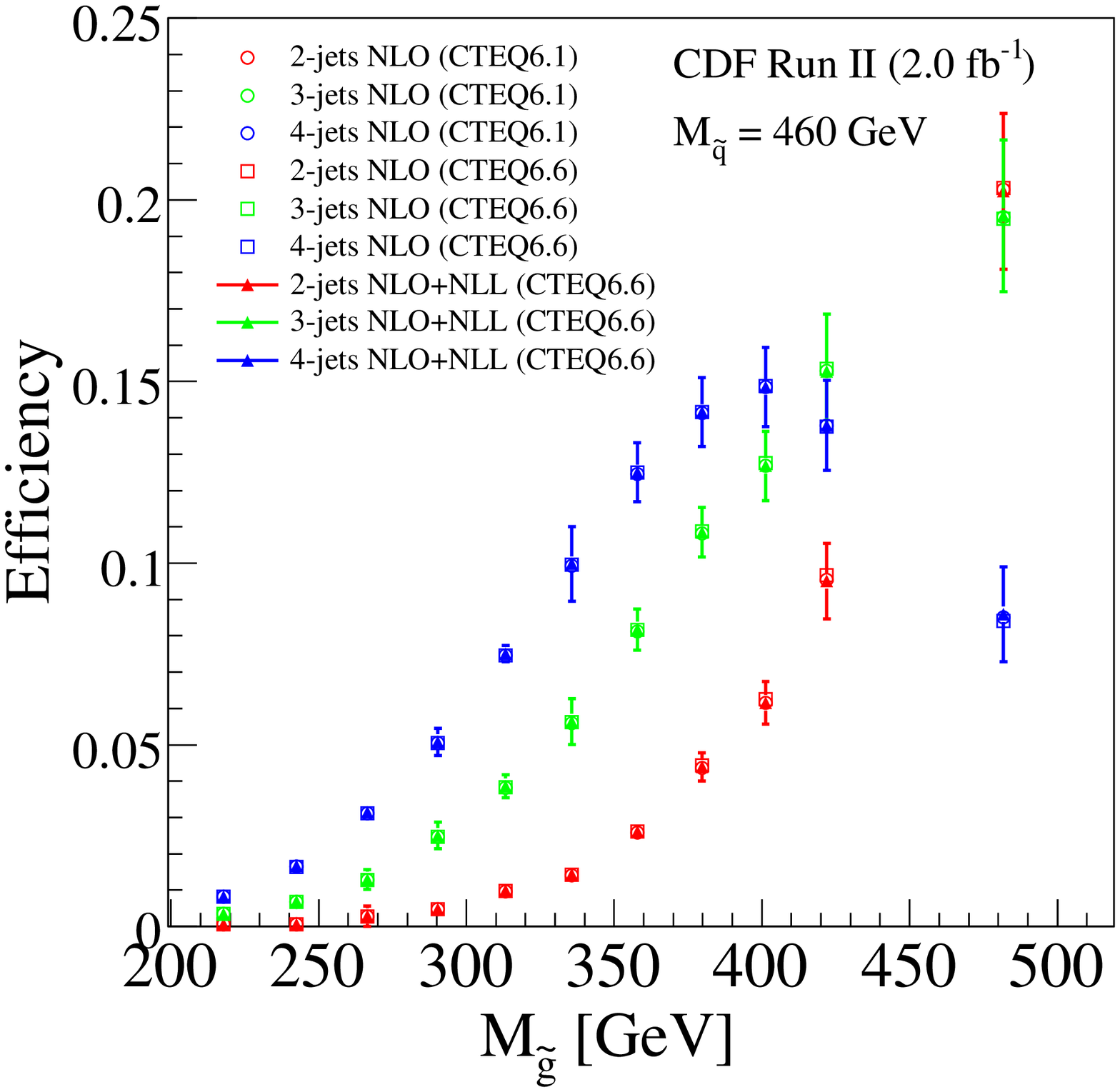, width=0.465\columnwidth }\\
\epsfig{file=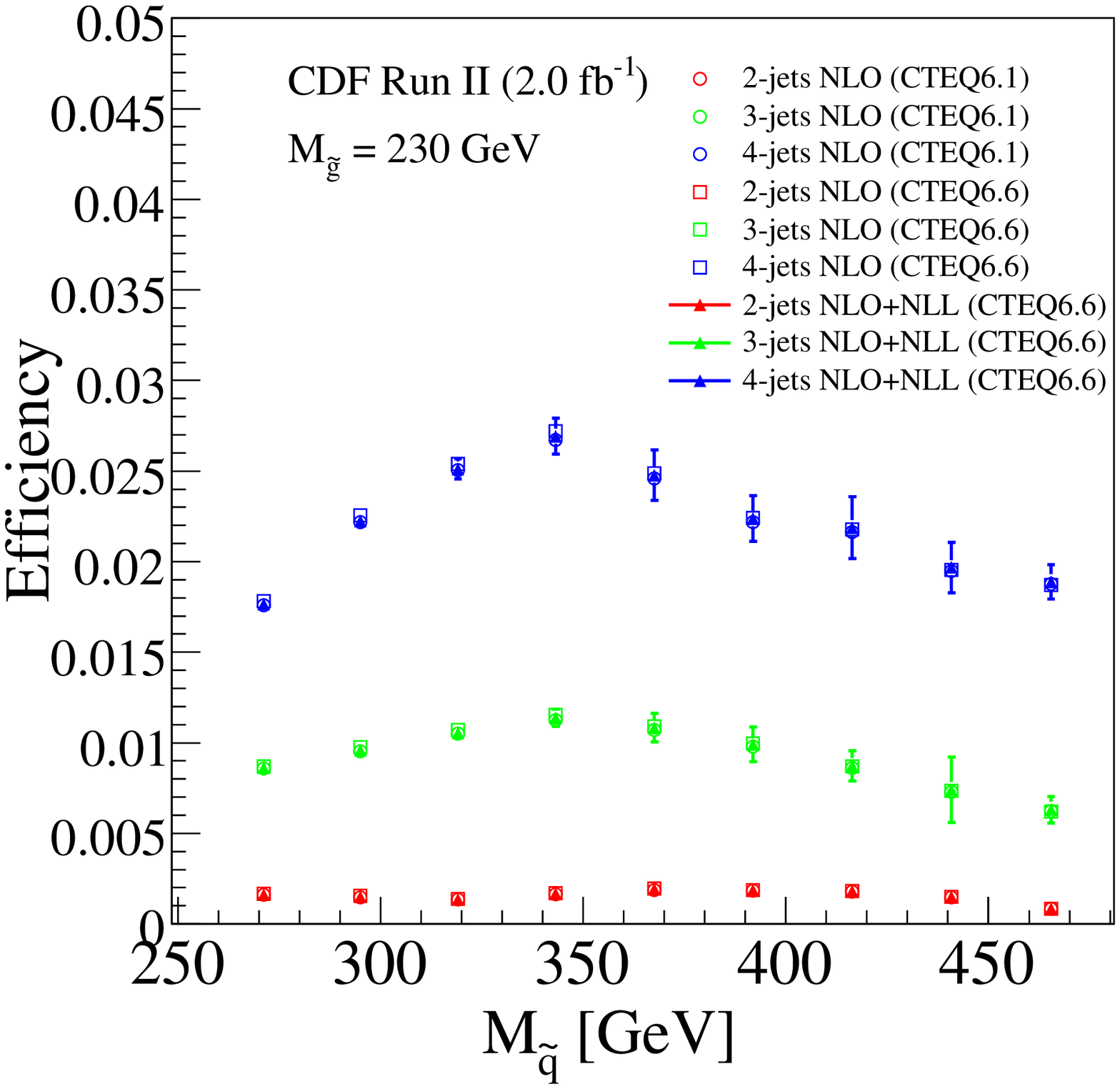, width=0.465\columnwidth}& 
 \\
\end{tabular}
\caption{
SUSY signal selection efficiencies for different inclusive jet multiplicities, as determined using PYTHIA MC samples normalized according to  NLO cross sections 
with CTEQ6.1M\,\cite{Stump:2003yu} and CTEQ6.6\,\cite{Nadolsky:2008zw} PDFs, or using NLO+NLL cross sections with CTEQ6.6 PDFs. 
Different squark and gluino mass configurations are considered and the efficiencies are 
presented as a function of squark and gluino masses.
}
\label{fig:effi}
\end{figure}

A variation in the theoretical predictions can potentially translate into a change in the 
signal selection efficiency if the mixture among subprocesses contributing to the same final state 
is modified significantly. 
The effect of updated PDFs and improved NLO+NLL predictions is considered
separately, and the variation of the signal selection efficiency is studied as a function of the squark and gluino
masses in three different configurations already considered by CDF~\cite{cdfweb}: degenerate squarks and gluinos, fixed squark mass $m_{\sq} = 460$~GeV, 
and fixed gluino mass $m_{\gl} = 230$~GeV.
Figure~\ref{fig:effi} shows the results 
separately for each jet multiplicity.
In all cases, the changes in the signal selection efficiencies are small, indicating that the 
updated theoretical predictions do not introduce a significant variation.

\begin{figure}
\hspace{-0.2cm}
\begin{tabular}{ll}
\epsfig{file=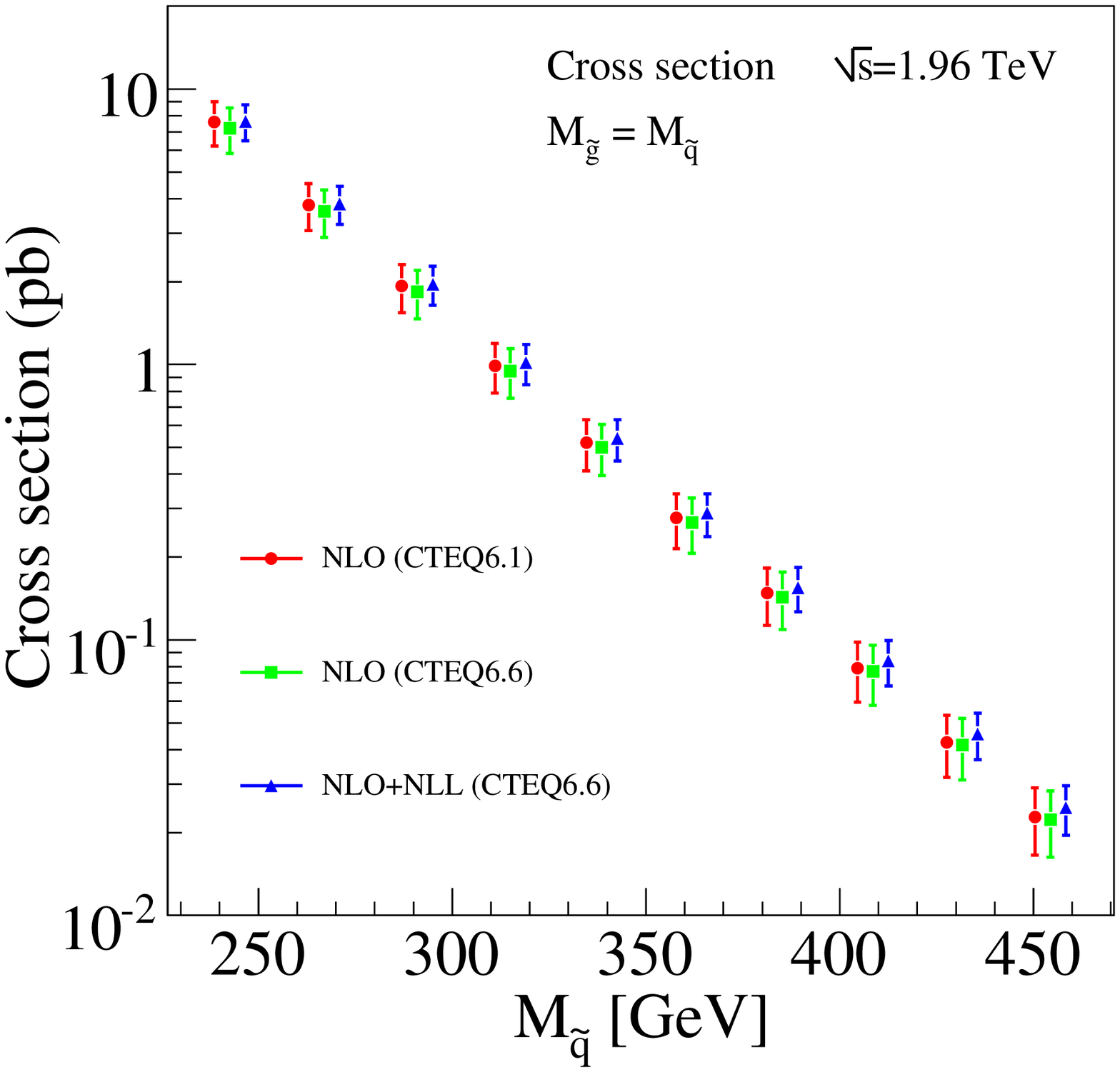, width=0.465\columnwidth}& 
\;\;\;\epsfig{file=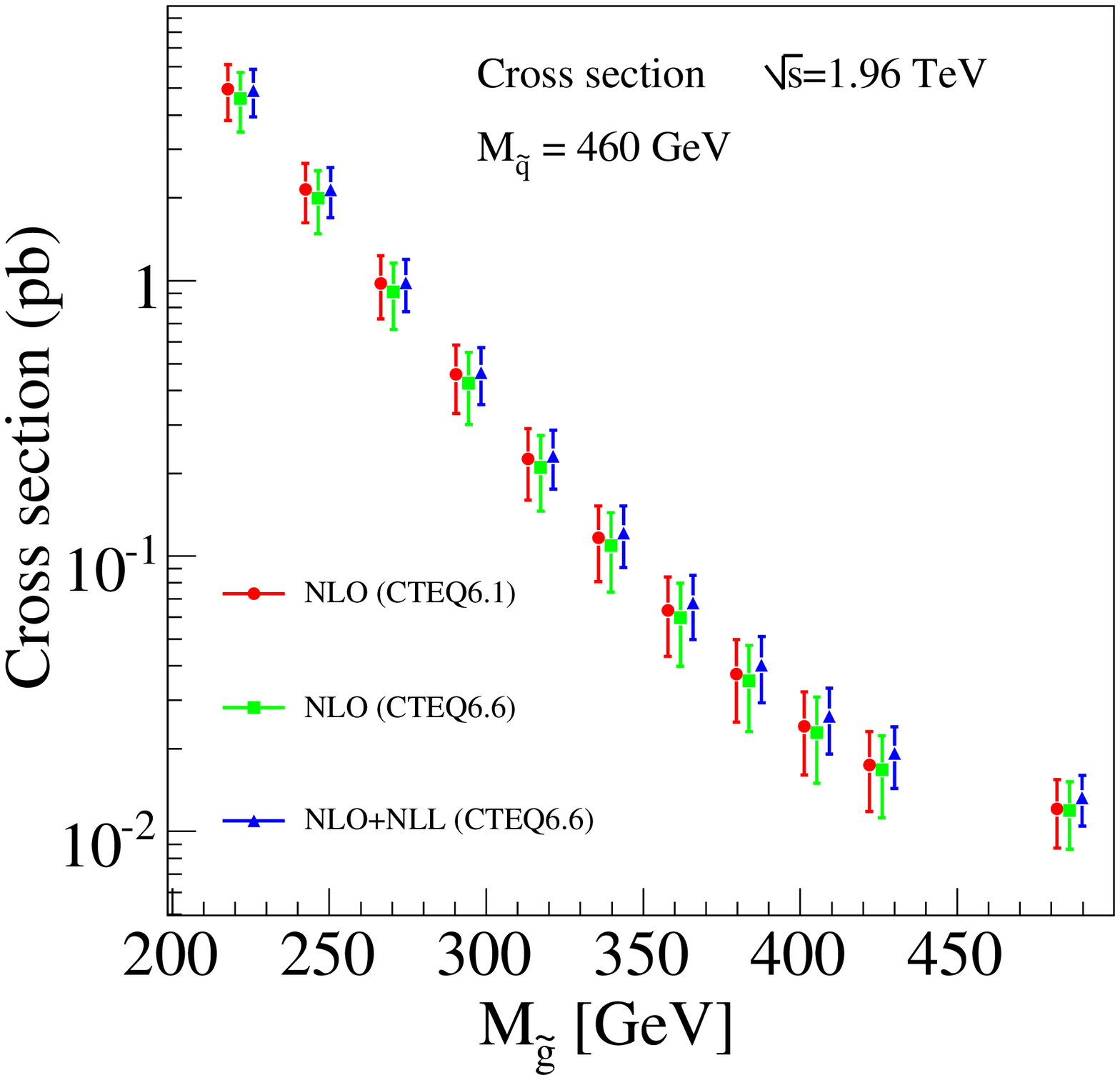, width=0.465\columnwidth }\\
\epsfig{file=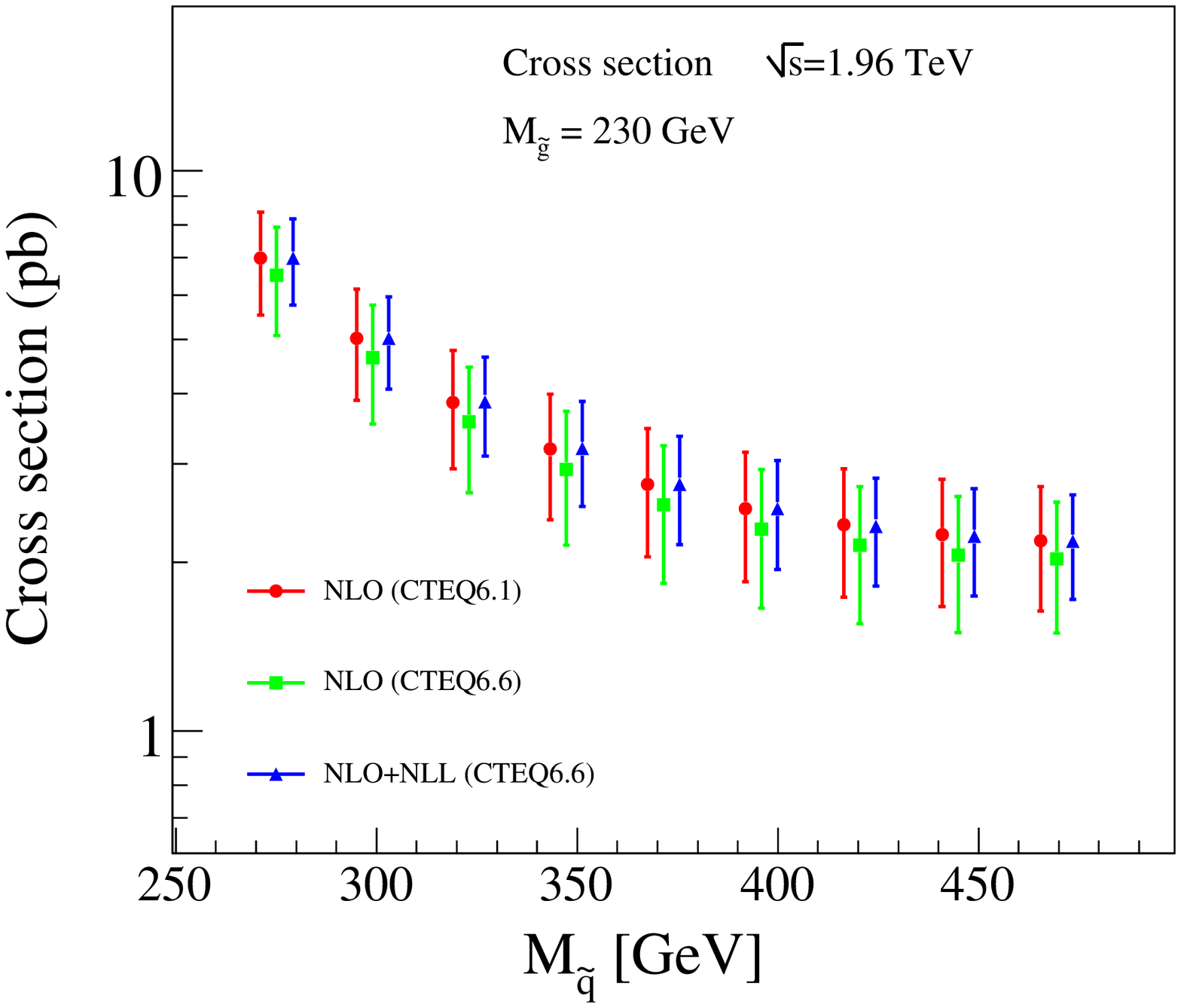, width=0.465\columnwidth}& 
 \\
\end{tabular}
\caption{
NLO cross section prediction using CTEQ6.1M\,\cite{Stump:2003yu} and CTEQ6.6\,\cite{Nadolsky:2008zw} PDFs, and 
NLO+NLL cross sections with CTEQ6.6 PDFs. 
Different squark and gluino mass configurations are considered and the predictions are 
presented as a function of squark and gluino masses. The error bars indicate the PDF and renormalization and 
factorization scales added in quadrature. Note that for display purposes the NLO and NLO+NLL CTEQ6.6 cross 
sections are shown at mass coordinates shifted by 4~GeV and 8~GeV, respectively, with respect to the NLO CTEQ6.1 
cross section.}
\label{fig:cross}
\end{figure}

The new predictions for the  SUSY cross sections in the different squark/gluino mass configurations considered above 
are compared to those used by CDF in Figure~\ref{fig:cross}. Again, the impact 
of the PDF set employed and the inclusion of the new  NLO+NLL terms are studied separately. 
In general,  the use of the CTEQ6.6 PDFs slightly reduces the theoretical cross sections, whereas the 
NLO+NLL calculations predict an enhanced SUSY signal cross section with a  reduced scale dependence. 
As a result, the new predictions differ from the previous ones by less than $1\%$ 
across the different squark and gluino masses considered, but present smaller uncertainties.   
Only in the scenario of similar squark and gluino masses, the new SUSY cross sections 
are significantly larger, and the difference varies from 0.3$\%$ to $10\%$ with 
increasing mass between 250~GeV and 460~GeV.   
       
\begin{figure}
\hspace{-0.2cm}
\mbox{
\epsfig{file=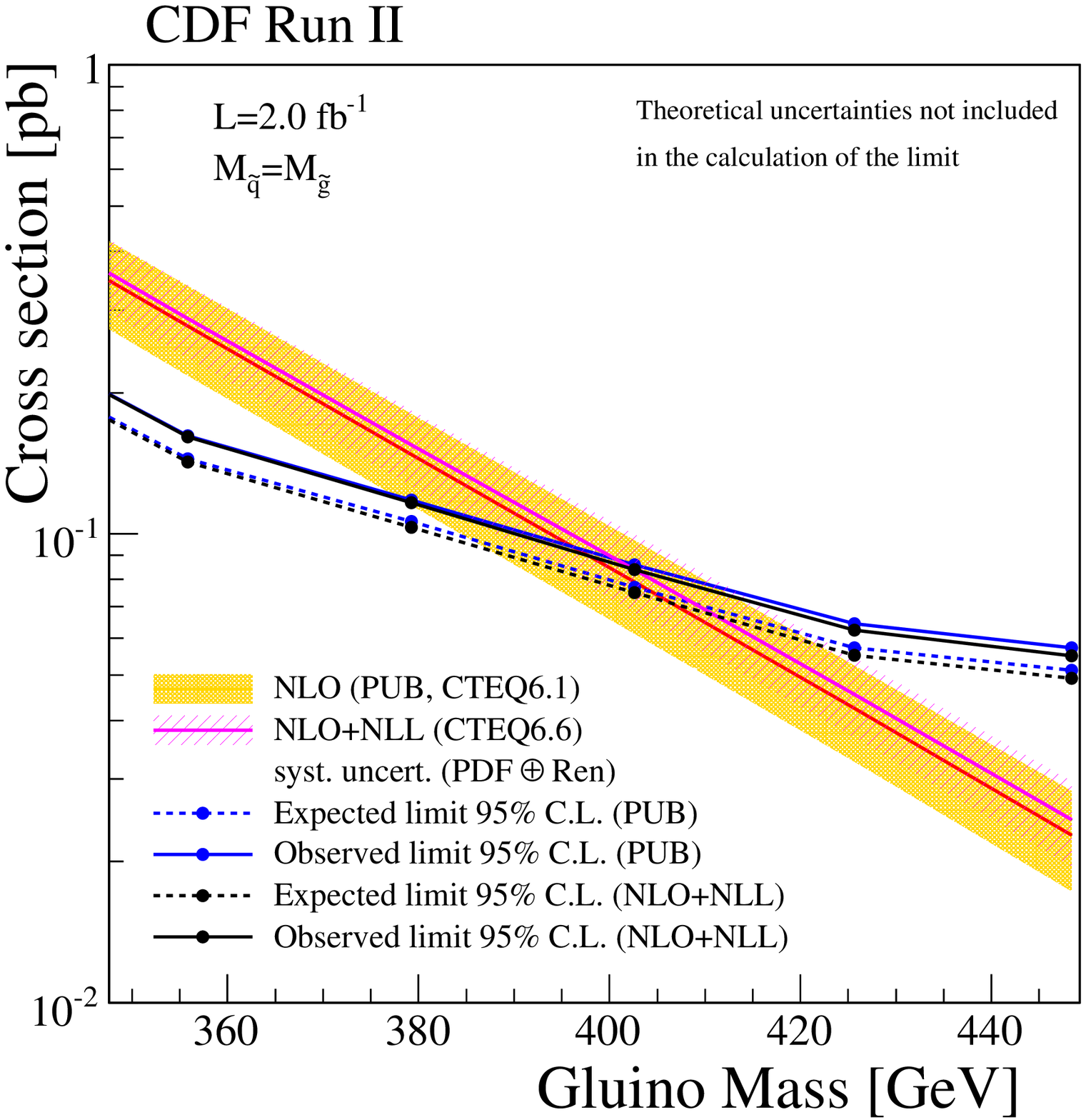, width=0.465\columnwidth} 
}
\caption{Expected and observed 95$\%$ C.L. upper limit on the 
SUSY production cross section in the case of degenerate squarks and gluinos.
The new limits (using NLO+NLL and CTEQ6.6 PDFs theoretical predictions) 
are compared to previous CDF published results~\cite{Aaltonen:2008rv}.   
}
\label{fig:limits_cross}
\end{figure}

\begin{figure}
\hspace{-0.2cm}
\mbox{
\epsfig{file=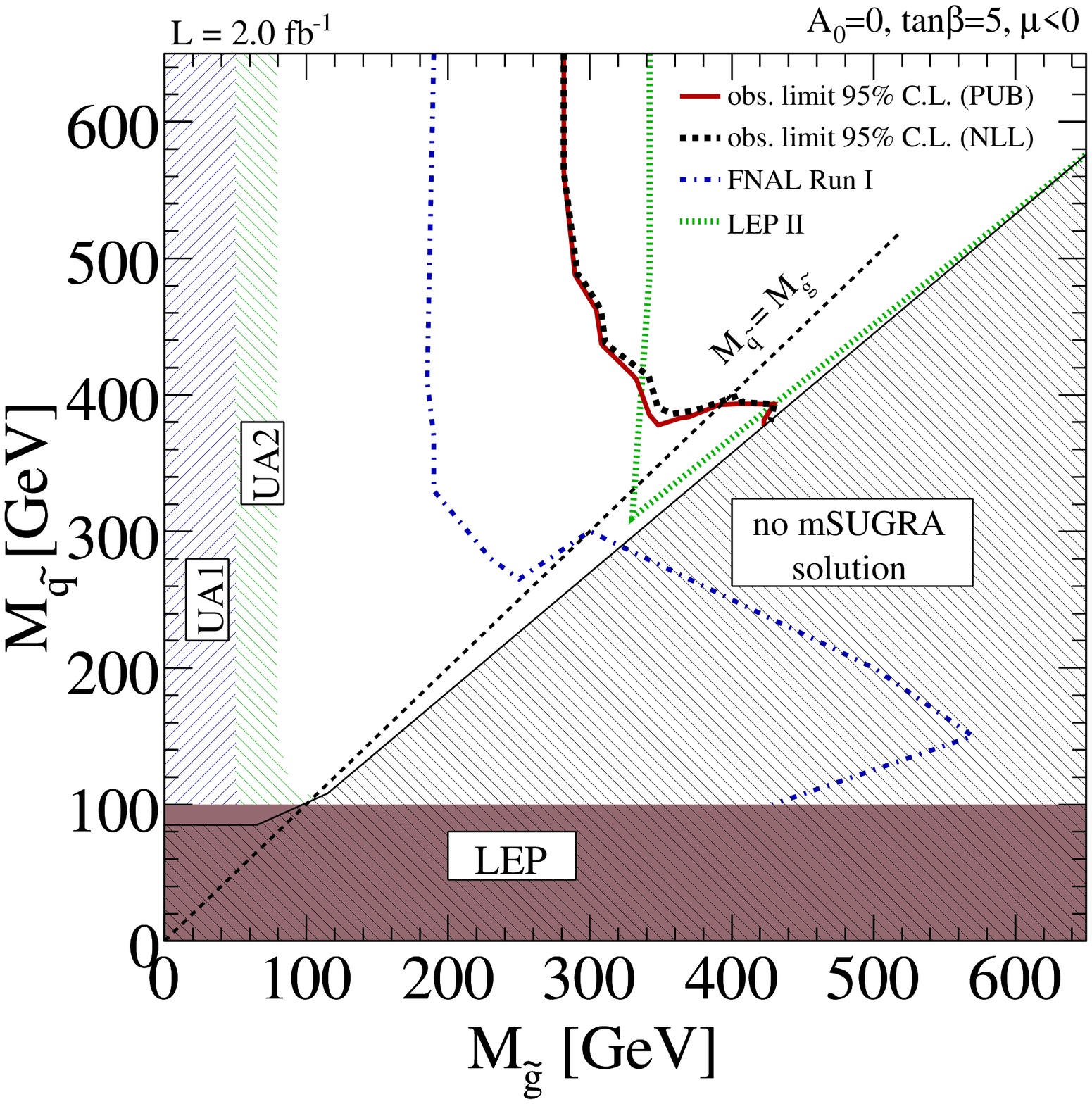, width=0.465\columnwidth}
}
\caption{Observed 95$\%$ C.L. exclusion region in the squark/gluino mass plane. The new results 
(dashed line) are compared to CDF published results~\cite{Aaltonen:2008rv} (solid line). 
}
\label{fig:limits_plane}
\end{figure}


\section{Squark and gluino experimental bounds}
\label{se:massbounds}

The effect of the new improved theoretical cross sections  on the 
calculation of the 95$\%$ confidence level (C.L.) upper limits for SUSY production 
cross sections and squark and gluino masses is investigated.   
In a counting experiment, the Poisson probability of observing $n$ events with an 
expected background $b$ and a signal efficiency $\epsilon$ is

\begin{equation}
\frac{e^{-(L s\epsilon + b)}(L s\epsilon + b)^n}{n !}, 
\end{equation}

\noindent
where $L$ is the integrated luminosity and $s$ is the signal cross section. 
Following the original CDF study, a 
 Bayesian approach~\cite{bayes} is adopted to compute the 95$\%$ C.L. upper limits on $s$ given the observed 
number of events in the data. The calculation
includes  statistical and systematic uncertainties on $\epsilon$ and $b$ and their correlations,  
and a $6 \%$ uncertainty on $L$.
In order to compute the mass bounds, the theoretical uncertainties on $s$ are also included in 
the limit calculation via an extra uncertaity on $\epsilon$. 
For each mSUGRA point considered, observed and expected limits are 
computed separately for each of the three 
analyses, corresponding to different jet multiplicities in the final state, and the one with the 
best expected limit is adopted as the nominal result.

Figure~\ref{fig:limits_cross} shows, in the case of degenerate squarks and gluinos,  
the new expected and observed 95$\%$ C.L. upper limits on the production cross section compared to 
previous results. As anticipated from  the observation of nearly invariant signal efficiencies after the inclusion 
of the  NLL terms in the calculation, the obtained cross section upper limits are close to those
published by CDF.  The results on the new squark and gluino mass bounds are 
presented in Figure~\ref{fig:limits_plane} for the whole squark/gluino mass plane. The 
impact from the new  theoretical predictions on the squark and gluino mass limits is  modest   
and only relevant for the case of similar squark and gluino masses, extending the CDF  excluded 
region by 5 to 10 GeV.

\section{Conclusions}
Precise theoretical predictions for sparticle cross sections are
essential for the interpretation of current and future searches for
supersymmetry at hadron colliders. Recently, the NLO-QCD 
cross section calculation has been further improved by the 
inclusion of threshold resummation at next-to-leading-logarithmic (NLL) 
accuracy. The NLL corrections 
reduce the renormalization- and factorization-scale dependence of the
predictions.  In general the NLL contributions also increase the cross 
section if the renormalization and factorization scales are chosen close to the
average mass of the pair-produced sparticles. 

The impact of the new improved theoretical NLO+NLL calculations on the experimental 
bounds for squark and gluino production has been shown using CDF results as 
benchmark. Improved 95$\%$ C.L. upper limits on the SUSY production cross section and 
the squark and gluino masses are presented in the case of nearly degenerate squarks 
and gluinos. The difference with respect to the CDF published bounds 
are found to be small once the new NLO+NLL cross section predictions are combined 
with updated CTEQ6.6 PDF sets.

\label{se:conclusion}

\section*{Acknowledgments}
\noindent 
The work has been supported by the Helmholtz
Alliance ``Physics at the Terascale'', the DFG Graduiertenkolleg
``Elementary Particle Physics at the TeV Scale'', the Foundation for
Fundamental Research of Matter (FOM) program "Theoretical Particle
Physics in the Era of the LHC", the National Organization for
Scientific Research (NWO), the DFG SFB/TR9 ``Computational Particle
Physics'', and the European Community's Marie-Curie Research Training
Network under contract MRTN-CT-2006-035505 ``Tools and Precision
Calculations for Physics Discoveries at Colliders''. MK would like to thank 
the Institute for High Energy Physics (IFAE) at the Universitat Aut\'{o}noma 
de Barcelona for hospitality and the Deutsche Forschungsgemeinschaft 
DFG for financial support. 

\bibliographystyle{h-physrev}

\end{document}